\documentclass[prl,twocolumn,amsmath,a4paper]{revtex4}

\usepackage{graphicx}

\begin{document}
\title{Measuring overlaps in mesoscopic spin glasses {\it via} conductance fluctuations}

\author{David Carpentier}
\affiliation{CNRS  UMR 5672 - Laboratoire de Physique de l'Ecole Normale
Sup{\'e}rieure de Lyon, \\
46, All{\'e}e d'Italie, 69007 Lyon, France}

\author{Edmond Orignac}
\affiliation{CNRS UMR 5672 - Laboratoire de Physique de l'Ecole Normale
Sup{\'e}rieure de Lyon, \\
46, All{\'e}e d'Italie, 69007 Lyon, France}

\date{23 Juillet 2007}

\begin{abstract}
We consider the electonic transport in a mesoscopic metallic spin glasses. 
We show that the distribution of overlaps between spin configurations
can be inferred from the reduction of the conductance fluctuations by the magnetic impurities. 
Using this property, we propose new experimental protocols to probe 
spin glasses directly through their overlaps. 
\end{abstract}

\maketitle

 Understanding the physics of glasses remains one of the deepest 
experimental and theoretical 
 challenge in condensed matter.
 Considered as the simplest glassy phases, 
  spin glasses have attracted considerable attention 
both experimental and theoretical during more than thirty years
\cite{lesHouches78,Parisi:2007}. 
  In spin glasses, magnetic
 moments occupying random positions in a host lattice get frozen with
 random orientation below
 a spin glass transition temperature $T_{SG}$. 
 The initial theoretical efforts were  devoted to the solution of 
equilibrium lattice models. This culminated with   
the mean-field solution of the fully connected Sherrington-Kirckpatrick (SK)
model \cite{Parisi:1980}, 
 now proven to provide the exact free energy \cite{Guerra:2002,Talagrand:2006}. 
 However, an intense theoretical debate remains around the 
relevance of the SK model 
to the thermodynamic properties of three dimensional spin glasses.  
 Although the initial theoretical studies were focused on the thermodynamics, 
 it was long known that spin glass materials  drop  out of
equilibrium below $T_{SG}$, and never reach a steady state. This has
led to the development of various models of non-equilibrium spin
glass dynamics including a scaling approach \cite{Bray:1986,Fisher:1986}, 
phenomenological trap models \cite{Bouchaud:1992}, and more recently 
aging studies of the mean field models 
(see \cite{Cugliandolo:2002} and ref. therein). 
 In the simplest scaling approach  to spin glasses, the thermodynamics below $T_{SG}$ 
 consists, in constrast to mean-field solutions,  
 of a doubly degenerate 
 broken $Z_2$ symmetry ground state  (for Ising spins). All
 the non-trivial properties of the spin glass state are then 
assumed  to be consequences of the 
 extremely slow relaxation toward this ground state resulting 
 from the slow growth of small domains (droplets) of equilibrium phase, 
 similar to ferromagnet quench. 

 In view of the current understanding of the spin glass state, it is a worthwhile goal 
to propose new probes of their properties. 
A central quantity in current theoretical descriptions of spin glasses is the 
overlap between two spin configurations defined as ($N_{imp}$ being 
the number of spins) 
\begin{equation}\label{def:overlap}
Q_{12} = \frac{1}{N_{imp}} \sum_{i=1}^{N_{imp}}
\langle \vec{S}^{(1)}_{i}. \vec{S}^{(2)}_{i} \rangle_{th} . 
\end{equation}
Configurations overlaps gives access to 
 distances between spins configurations, as opposed to conventional probes like 
 spin susceptibility.
 For configurations corresponding to equilibrium states in the same sample, 
 overlap distribution is the equilibrium order parameter \cite{Parisi:2007}. For configurations 
 taken in the same quench but at different times, overlaps characterize the  
 glass aging \cite{Cugliandolo:2002}. Besides, other physical effects such as
  temperature or disorder chaos \cite{Katzgraber:2007}
 are described in terms of spins overlaps.
 It is the purpose of this letter to propose 
 the first experimental probe of these central quantities
{\it via } the study of conductance fluctuations in mesoscopic metallic 
spin glasses.

 The dependance of the average conductance fluctuations on magnetic
 impurities was first analyzed by Althuler and Spivak
 \cite{Altshuler:1985}. Soon after, Feng {\it et al.} \cite{Feng:1987}, 
  building on these results, predicted {\it within the scaling
    approach} a  chaotic behavior of conductance as a function of
  temperature in a spin glass. Parallel to these theoretical
  developments, experimental measurements of conductance fluctuations 
in metallic spin glasses  by de Vegvar {\it et al.}
 \cite{deVegvar:1991} (see also \cite{Israeloff:1989}) 
 demonstrated for the first time a clear
 signature of the spin glass freezing in the time-reversal
 antisymmetric part of the four terminal conductance.  Later, 
several experiments focused on noise measurements in Cu:Mn    \cite{Alers:1992}
and Au:Fe  \cite{Neuttiens:2000} (see also \cite{Jaoszynski:1998} for 
 similar studies in the  doped semiconductor). 
%
 
 However, a connection between these 
 theoretical and experimental analyses and spin glass theories was difficult as none of these 
 approaches allowed a simple interpretation within the spin glass theoretical framework. 
  This motivates a  reexamination of the description of conductance fluctuations in a spin glass 
 in relation with spin overlaps. 

 We consider a mesoscopic metallic sample of size $L$ containing magnetic impurities. 
 These impurities provide three different contributions to the scattering potential for the conduction electrons : (i) a scalar potential 
 $V({\bf r}) = \sum_{i} v_{o}\delta ({\bf r}-{\bf r}_{i})$ where ${\bf r}_{i}$ denotes the positions of impurities, 
(ii) a spin coupling  $V_{S}= J(T) \sum_{i=1} \vec{S}_{i}.\vec{\sigma}_{e}$,  and (iii) a spin-orbit 
contribution 
$V_{so}(\vec{k}_{1},\vec{k}_{2})= i V_{so} (\hat{k}_{1} \times \hat{k}_{2} ).\vec{\sigma}_{s_{1}s_{2}}$ . 
 The magnetic impurities interact with each other {\it via} a RKKY interaction  \cite{Mydosh:1993}
 $\sum_{i\neq j}J_{ij}~\vec{S}_{i}.\vec{S}_{j} $, 
 and at high enough impurity concentrations, the corresponding spin
 glass transition temperature $T_{SG}$  is larger  
the Kondo temperature $T_K$.  In this regime, the local moments remain 
unscreened at $T_{SG}$ and as a result,  due to the random 
signs of the couplings,  they freeze into the spin glass
state for $T<T_{SG}$ \cite{Mydosh:1993}. 
In the spin glass state, the frozen spins act on the conduction
electrons  just like a classical random magnetic field. In the rest of
the letter, we will  focus on this regime ($T_{SG} >T \gg T_{K}$) where
transport properties simply result from coherent diffusion of
electrons by both a  classical random magnetic field and the associated 
scalar potential.

 Let us start by recalling known results about the fluctuations of conductance induced by a scalar 
 random potential $V({\bf r})$. We denote by $L_{\phi}$ the dephasing or inelastic 
 scattering length, which is considered larger than the sample size $L$ (mesoscopic regime). The associated inelastic scattering 
 rate is $\gamma_{\phi} = \hbar / (2 \tau_{\phi})= D \hbar /(2 L_{\phi}^2)$, D being the diffusion constant 
 in the sample. 
 For the sake of clarity, we focus on the longitudinal $G=G_{xx}$, although the
following discussion  extends naturally to other components of the conductance \cite{footnote:Gxy}.
In this mesoscopic regime, the conductance $G$ is a function of the random scattering potential $V({\bf r})$, and in the weak disorder limit, 
its distribution is approximately gaussian (see \cite{Tsyplyatyev:2003} 
for a recent discussion). Its average 
incorporates  weak-localization corrections \cite{Hikami:1979,Wei:1988}, and 
 its variance, describing the sample to sample fluctuations, contains contributions from both fluctuations of 
 the diffusion coefficient and of the density of states (see \cite{Akkermans:2007} for a pedagogical introduction). 
 With  only a scalar potential $V({\bf r})$ and in the so-called diffusion limit, 
 the fluctuations of this conductance read for  weak disorder : 
\begin{equation}\label{eq:variance-G}
\langle (\delta G )^2 \rangle_{V}   = F (\gamma_{\phi}) = 
 6 \left( \frac{e^2 D}{h L^2} \right)^2  \sum_{\vec{q}}  
\left( D q^2  + \gamma_{\phi} \right)^{-2}
\end{equation}
where $\delta G  = G  -\langle G \rangle_{V} $. 
For a wire where diffusion takes place in one-dimension (1D) between two absorbing reservoirs, the variance (\ref{eq:variance-G}) 
reduces to $\langle (\delta G )^2 \rangle_{V}  = 8/15 (e^2/h)^2$, the so-called 
universal conductance fluctuations in the limit $L\ll L_{\phi}$ . In the other case $L_{\phi} < L$, these 
fluctutations reduce 
to 
$\langle (\delta G )^2 \rangle_{V}  \simeq (e^2/h)^2 (L_{\phi}/L)^{4-d} $ with a geometrical factor. 

 How are these results modified in the presence of the random field component $V_{S}$ of the scattering potential ?
First, those spins that can flip during the electron 
diffusion time (either weakly connected or maximally frustrated)
 will contribute to the enhancement of the inelastic scattering rate $\gamma_{\phi}$. 
We assume that the inelastic coherence 
length of the sample, including the effects of these quasi-free spins, is still larger than the system size at
 low enough temperatures. The remaining spins are considered as classical random fields, frozen on the electrons 
diffusion time-scale. These random fields ``flip'' the electron spin, and thus provide a finite lifetime to 
different diffusion spin states. 
Using a semi-classical approach allows us to consider a given realization of spins, without averaging. 
A diffusion path is labeled  by the sequence of encountered 
impurities, ordered chronologically. At each impurity $j$, the electron's spin is rotated according to  
$
\mathcal{R}_{j} = e^{i J \vec{S}_{j}.\vec{\sigma}} = \cos(JS) +i \sin(JS) \hat{S}.\vec{\sigma} .
$
 The end action of the random fields along the path is encoded by the chronological product 
  $\prod_{j} \mathcal{R}_{j}$. 
Expanding this product 
 in the limit of weak  $J$, and using the central-limit theorem
 we obtain the {\it typical} magnetic dephasing rate of an electron state as 
$
\gamma_{m} =  2 \pi \rho_{0} ~ n_{imp} J^2 \langle S^2 \rangle_{th} 
$, 
 where $\rho_{0}$ is the density of state, $n_{imp}$ the concentration of impurities and 
 $\langle \rangle_{th}$ means an average over thermal fluctuations. 
 Note that in doing so, we have neglected all spatial spins correlations, a coherent assumption in a spin glass 
state. Moreover, we have assumed 
a good impurities sampling by typical diffusion path, approximating the number of impurities along typical path 
by $N_{imp}$.  

 Coming back to the conductance fluctuations, we assume that both sources of 
 disorder  $V({\bf r})$ and $V_{S}({\bf r})$ can be treated as 
independent from each other (the
orientation of the frozen moment is not directly correlated with the
position of the single impurity).  
 We focus on the average (over $V$) correlations 
between conductances in a given sample $V({\bf r})$
with two different spin configurations $\{S_{j}^{(1)} \}$ and $\{S_{j}^{(2)} \}$  
 \begin{equation}\label{def-GG}
 (\Delta G)^2_{S^{(1)}S^{(2)}} = 
\left< \delta  G  \left(V,\{S_{j}^{(1)} \}\right)  \delta G\left(V,\{S_{j}^{(2)} \}\right) \right>_{V} 
\end{equation}
 where 
 $\delta  G  (V,\{S_{j}^{(1)} \})  = G  (V,\{S_{j}^{(1)} \})  - \langle G  (V,\{S_{j} \}) \rangle_{V}$. 
 The weak disorder expression of this average correlation is obtained 
 similarly to (\ref{eq:variance-G}) using standard diagrammatic techniques 
  (see \cite{Akkermans:2007}). 
 In doing so, one formally considers the diffusion of the so-called Cooperons and 
 Diffusons. 
 They can be viewed as the coherent propagation of pairs of electrons along a path in 
 the same chronological order (Diffuson), of in reversed orders (Cooperon). In the correlations 
 (\ref{def-GG}) the two components of the Diffuson or Cooperon 
  see  {\it a different spin configuration}. The action of the magnetic impurities on their diffusion is 
\begin{equation}\label{eq:rotations-Cooperon}
\prod_{j} e^{i (JS) \hat{S}^{(1)} _{j}.\vec{\sigma}^{(1)}} e^{\pm i (JS) \hat{S}^{(2)} _{j}.\vec{\sigma}^{(2)}}, 
\end{equation}
 where the $\pm$ sign depends on the nature of the diffusive object.  The Diffuson/Cooperon states cary a pair of spins $1/2$, and are naturally decomposed 
into a singlet and triplet states. 
 In a given spin configuration
($\vec{S}_{j}^{(1)}=\vec{S}_{j}^{(2)}$), only the triplet states couple to these 
random magnetic fields while for two different 
spin configurations $\{S_{j}^{(1)} \},\{S_{j}^{(2)} \}$, both the singlet and the triplet states 
 acquire a finite diffusion lifetime. From eq.~(\ref{eq:rotations-Cooperon}), we obtain the 
 {\it typical} dephasing rate of these composite Diffusons in the limit of weak $J$ :
$\gamma_{m}^{D,S}   =  \gamma_{m} (1-Q_{12}),
\gamma_{m}^{D,T}   =  \gamma_{m} (1+\frac13 Q_{12})
$
 where $Q_{12}$ is defined in (\ref{def:overlap}). 
 The scattering rates for the Cooperons follow by $Q_{12}\to -Q_{12}$. 
  Now plugging these scattering rates back into the diffusion propagators,  
we obtain the following expression for the average correlations :
\begin{align}
\nonumber
 (\Delta G)^2 &= 
\frac14 F \left(\gamma_{m}^{D,S}+\gamma_{\phi} \right) 
+ \frac34 F  \left(\gamma_{m}^{D,T}+ \gamma_{so}+\gamma_{\phi}\right)
\\
&+ \frac14 F \left(\gamma_{m}^{C,S}+\gamma_{\phi} \right) 
+ \frac34 F  \left(\gamma_{m}^{C,T}+ \gamma_{so}+\gamma_{\phi}\right) 
\label{eq:variance-G-spins}
\end{align}
where we have included the spin-orbit and inelastic dephasing rates. 
When (some) magnetic dephasing lengths are smaller than 
$L$ ($ < L_{\phi},L_{so}$), 
the fluctuations are dominated by the smallest dephasing rate (one of the singlets). 
For $\gamma_{\phi}\ll E_{c} \ll \gamma_{m} (1-Q_{12})$, 
we obtain  for $D=2$ : 
\begin{equation}\label{eq:approx-GG}
\left< \delta  G  (\{S_{j}^{(1)} \})  \delta G(\{S_{j}^{(2)} \}) \right>_{V} \propto 
\left(\frac{e^2}{h} \right)^2 \frac{E_{c} }{\gamma_{m} (1-Q_{12})}   . 
\end{equation}
where $E_{c}$ is the Thouless energy.  
Crucially, 
these correlations depends on the quantity $Q_{12}$, which is called
the spin {\it overlap} and plays a central role in  the 
description of spin glasses. 
Indeed, the  distribution of $Q_{12}$ is the  spin glass order parameter in the mean-field theory.    
 
\begin{figure}[!t]
\centerline{\includegraphics[width=7cm]{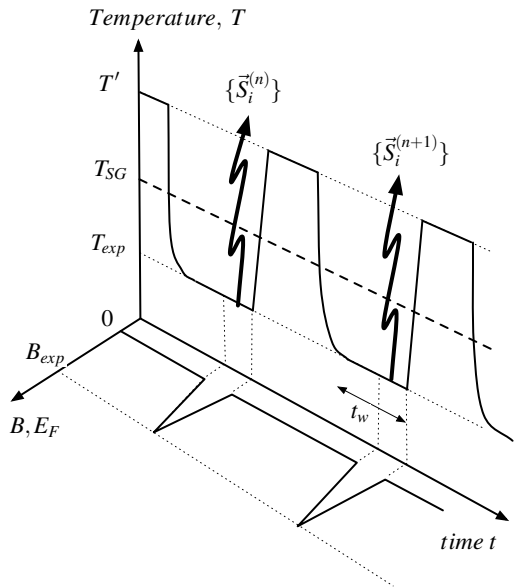}}
\caption{Proposed experimental protocol : the temperature is cycled through $T_{g}$. 
Measurements at successive reach $n$ of temperature $T_{exp}$ correspond 
to different spins states $\{\vec{S}_{i}^{(n)}\}$ in a given sample, labelled by an index $n$. 
The correlations $\langle \delta G_{n} \delta G_{m} \rangle_{B}$ between these spin configurations, 
and the corresponding  overlap $Q_{nm}$ are obtained by application of small magnetic fields after 
some time $t_{w}$. 
 Repeating this process gives access to the (non-average) overlap distribution for
different waiting time $t_{w}$.
}
\label{fig:temp-cycling}
\end{figure}
Before pursuing further into the spin glass considerations, 
we first discuss the disorder averaging in experiments. 
In the above analysis, we have carefully avoided to average over
spin configurations while at the same time  averaging over the scalar
potential. However, in experiments, both disorder originates from the
same source {\it i.e.} the random positions of magnetic impurities.
We thus need to propose an experimental setup that 
simulates an average over a scalar potential while keeping the 
 distribution of spins (quasi-) fixed.  
Experiments usually rely on the ergodic hypothesis and probe these 
fluctuations by varying the magnetic field or the Fermi energy. 
Physically the origin of conductance fluctutations are phase-coherent
 contributions, encoded into the Cooperon and Diffuson. 
A change in disorder changes the various 
diffusion paths, and the corresponding interferences. However, the
various phases can  also be modified  either by 
the orbital effect of a uniform  magnetic field or a 
change of the Fermi energy, hence the ergodic hypothesis. 
Averaging over the Fermi energy can be achieved in  doped semi-conductor spin
glasses by applying a gate potential \cite{Jaoszynski:1998,footnote:EF-average} . 
In metallic spin glasses, one has to resort 
to the magnetic field sampling and the average in eq.~(\ref{def-GG}) is replaced 
by  $ \langle \delta G  (V,\{S_{j}^{(1)}\})  \delta G(V,\{S_{j}^{(2)}\})\rangle_{B}$ defined by 
$$
\frac{1}{B_{max}-B_{\phi}} \int_{B_{\phi}}^{B_{max}} 
 \delta G  (V,\{S_{j}^{(1)}\},B)  \delta G(V,\{S_{j}^{(2)}\},B) dB 
$$
The decorrelation field $B_{\phi}$ corresponds to two flux quanta
through the sample, and for the variance of the
conductance, relatively weak field amplitude $B_{max}/B_{\phi}$ are
necessary for a correct sampling  \cite{Tsyplyatyev:2003}. 
Note that
with a magnetic field, Cooperons are dephased, and
eq.~(\ref{eq:variance-G-spins}) simplifies to its first  line. 
 A crucial step for spins glasses, is to be able to find a magnetic field 
 $B_{max}$ at low enough temperature such that conductance fluctuations
 are enough sampled, while 
at the same time the magnetic response of the spins {\it can be neglected}. 
 This SG response effect can be experimentally estimated by applying an 
in-plane magnetic field, for which orbital effects can be neglected at small fields. 
  According to eq.~(\ref{eq:approx-GG}), the perturbation
of the spin configuration should manifests in the conductance fluctuations through 
the quantity $Q(\Delta B) =
N_{imp}^{-1} \sum_{i=1}^{N_{imp}} \vec{S}_{i}(B).\vec{S}_{i}(B+\Delta B)$. 
 The necessary condition for the proposed method is $Q(B_{max}) \simeq 1$ with probability of order $1$. 
 Although spin glasses are generally believed to show a chaotic magnetic response, 
to our knowledge current theoretical studies have focused on mean-field models with Ising spins, 
under fields larger than $0.1 T_{SG}$ \cite{Billoire:2003}, not directly 
relevant in the present context. 

Having access to overlaps of the spin configurations 
 opens new perspectives in probing the metallic spin glasses that we illustrate here by discussing three 
 possible experimental protocols. 
A)  the distribution of overlaps between spin configurations corresponding to different quench in the same sample 
 can be obtained from the previous ideas, as described in  Fig.~\ref{fig:temp-cycling}. In the limit of long 
 times $t_{w}$, this distribution provides information on the phase space structure of the spin glass. 
Obtaining this quantity for the first time in an experimental spin
glass would be of major importance as current theoretical proposals differ on its expected behavior. 
B)
Direct access to spin overlaps at different times allows for an unprecedented analysis of the aging of experimental 
spin glasses \cite{Cugliandolo:2002}. In a canonical experimental scheme, the sample is cooled down below $T_{SG}$ under  
a small magnetic field. This field is kept constant for a time $t_{w}$ and then switch off.  
Magnetic field sweeps provide  
$\langle \delta G (t_{w}) \delta G (t_{w}+t) \rangle_{B}$ and thus the overlap  
$Q(t_{w},t_{w}+t)$. By repeated cool down, both the average and crucially the statistics of this quantity can be determined. 
C) a procedure similar to A) allows to probe the temperature chaos of the spin glass \cite{Feng:1987}, and its relation with 
the rejuvenation ($T'<T_{exp}$) and memory ($T'>T_{exp}$) phenomena \cite{Bouchaud:2001}. 
The sample is kept at $T_{exp}$ for a time $t_{w}$ before applying a small field sweep. The temperature is then 
switched to $T'=T\pm \Delta T$  which now remains below $T_{SG}$. Successive magnetic sweep at times 
$t_{n}=t_{w}+n\tau$ are then applied at $T'$. Conductance correlations provides the overlap 
$Q(T_{exp},t_{w};T',t_{n})$ which characterizes  
the temperature chaos of the spin glass, and determine in particular the 
dependance of the overlap length $L_{c}$ on both temperature variation $\Delta T$ and time $t_{n}$. 
 The behavior of $L_{c}$ in relation with predictions from the droplet theory 
 are currently theoretically investigated  (see {\it e.g.} \cite{Bouchaud:2001,Katzgraber:2007}). 
 The goal is a characterization of low energy excitations of a spin glass state, and their slow evolution. 

  Let us end by commenting on the dimensionality of a mesoscopic spin glass. 
 To remain in the coherent transport regime, sample size $L$ 
 has to be of the order of (or lower than) the inelastic coherence length $L_{\phi}$ \cite{footnote:L-phi} 
 Often 
 quantum diffusion takes place in an effective space of dimension $D=1$. 
 On the other hand, the dimensionality of the spin glass is determined by the 
  dynamical correlation length. Values of the correlation length $\xi_{SG}=N_{SG}. n_{imp}^{1/3}$ ($n_{imp}$ 
  the density of spins)  
  extracted from field change experiments for various spin 
  glasses  \cite{Bert:2004} and extrapolation from recent numerical simulations \cite{Jimenez:2005} 
  are of the order of $N_{SG} \sim 30-50$ spins after a waiting time $t_{w}=1000 s$. 
  For samples with transverse dimensions $L_{y},L_{z}$ larger than 
$\xi_{SG}$, 
  the proposed conductance measurements will  probe properties of an effective 3D spin glass.  Reconsidering the experiments
  of \cite{deVegvar:1991}, we obtain approximately $\approx 40$ spins in the transverse dimensions  $L_{y}\simeq 900$~\AA, implying a 3D spin glass behavior. Moreover, this opens the perspective of studying a possible 3D to 1D crossover for the spin glass dynamics, and possibly determining the associated dynamical  correlation length. 
  
 To conclude, we have shown how new experimental protocols to determine mesoscopic conductance fluctuations in spin-glasses  can provide access to the spin configurations overlaps. We believe that these protocols can open completely new ways of characterizing the spin glass physics, and allowing progress in their understanding.

We aknowledge fruitful and motivating discussions with C. B\"auerle, F. Krzakala, L. L\'evy, L. Saminadayar and B. Spivak, and G. Montambaux for a correspondance regarding formula (\ref{eq:variance-G-spins}). 
 This work was supported by the  ANR Mesoglass.


\begin{thebibliography}{29}
\bibitem[{foo({\natexlab{b}})}]{footnote:EF-average}
\bibinfo{note}{Note that modulation of the electronic density can also modify
  the couplings between the spins. This effect will limit the range of gate
  potential usable to perform the average.}
\end{thebibliography}

\end{document}